\documentclass[a4paper,final]{appolb}
\usepackage{epsfig}
\usepackage{subfigure}
\begin{document}
\pagestyle{plain}
\title{Cylindrically asymmetric hydrodynamic equations.
  \thanks{Talk presented at the XLVI Cracow School of Theoretical Physics}
}
\author{Miko{\l }aj Chojnacki
  \address{The H. Niewodnicza\'nski Institute of Nuclear Physics,\\
    Polish Academy of Sciences,\\
    ul. Radzikowskiego 152, PL-31342 Krak\'ow, Poland}
}
\maketitle
\begin{abstract}
We show that the boost-invariant and cylindrically asymmetric hydrodynamic equations for baryon-free matter may be rewritten as only two coupled partial differential equations. In the case where the system exhibits the cross-over phase transition, the standard numerical methods may be applied to solve these equations. An example of our results describing non-central gold on gold collisions at RHIC energies is presented.
\end{abstract}
\PACS{25.75.-q, 25.75.Dw, 25.75.Ld, 25.75.Nq}
\section{Introduction}
\label{sect:Intro}
The success of the relativistic hydrodynamics in describing the RHIC data \cite{Teaney:2001av,Huovinen:2001cy,Huovinen:2002fp,Kolb:2002ve,Hirano:2002ds,Hirano:2004rs,Hirano:2005wx} suggests that the hot and dense matter produced at RHIC behaves like an almost perfect fluid \cite{Heinz:2005zg}. In this paper we discuss how the boost-invariant and cylindrically asymmetric relativistic hydrodynamic equations for baryon-free perfect fluid may be very conveniently reduced to only two coupled partial differential equations. Our presentation is based on the recent investigations published in Refs. \cite{Chojnacki:2004ec,Chojnacki:2006tv}. We argue that the effects of the cross-over phase transition may be included by the use of the temperature dependent sound velocity $c_s(T)$. As long as the function $c_s(T)$ satisfies the stability condition against the shock formation, the resulting equations may be solved with the help of standard numerical methods and used to describe the expansion of matter produced in the central region of ultra-relativistic heavy-ion collisions, such as studied in the present RHIC or future LHC experiments. 
The presented formalism is a direct generalization of the approach introduced by Baym, Friman, Blaizot, Soyeur, and Czyz \cite{Baym:1983sr} where the boost-invariant and cylindrically symmetric systems were considered, and the numerical calculations were performed only for the case of the constant sound velocity. 
\section{Hydrodynamic equations}
\label{sect:HE}
\begin{figure}[t]
\begin{center}
\includegraphics[angle=0,width=0.3\textwidth]{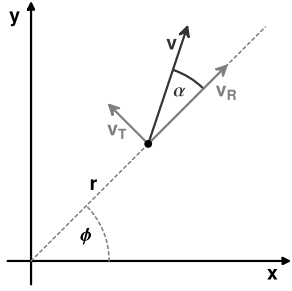}
\end{center}
\caption{Decomposition of the flow velocity vector in the plane $z=0$. In our approach we use the magnitude of the flow $v$ and the angle $\alpha$ as two independent quantities.}
\label{fig:alpha}
\end{figure}

The relativistic hydrodynamic equations of the perfect fluid follow from the energy-momentum conservation law and the assumption of the local thermal equilibrium. For baryon-free matter they have the following form
\begin{eqnarray}
u^\mu \partial_\mu (T u^\nu) = \partial^\nu T, \quad \quad \partial_\mu (s u^\mu) = 0, 
\label{h1} 
\end{eqnarray}
where $T$ is the temperature, $s$ is the entropy density, and $u^\mu = \gamma(1,{\bf v})$ is the four-velocity of the fluid.  For boost-invariant systems Eqs. (\ref{h1}) may be reduced to three independent equations \cite{Dyrek:1984xz}
\begin{eqletters}
\label{hyd2}
\begin{eqnarray}
\frac{\partial }{\partial t}\left( rts\gamma \right) +\frac{\partial }{\partial r}\left( rts\gamma v\cos \alpha \right) + \frac{\partial }{\partial\phi }\left( ts\gamma v\sin \alpha \right) &=&0, \label{h3} \\
\frac{\partial }{\partial t}\left( rT\gamma v\right) + r\cos \alpha \frac{\partial }{\partial r}\left( T\gamma \right) +\sin \alpha \frac{\partial }{\partial \phi }\left( T\gamma \right)  &=&0, \label{h4} \\
T\gamma ^{2}v\left( \frac{d\alpha }{dt}+\frac{v\sin \alpha }{r}\right) -\sin \alpha \frac{\partial T}{\partial r} +\frac{\cos \alpha }{r}\frac{\partial T}{\partial \phi } &=&0. \label{h5}
\end{eqnarray}
\end{eqletters}
In Eqs. (\ref{h3}) - (\ref{h5})  the quantities $t, r\!\!=\!\!\sqrt{x^2+y^2}$, and $\phi=\hbox{tan}^{-1} (y/x)$ are space-time coordinates which parameterize the plane $z=0$. The quantity $v$ is the magnitude of the fluid velocity, $\gamma =\left(1-v^2\right)^{-\frac{1}{2}}$ is the Lorentz factor, and $\alpha$ is the function describing direction of the flow, $\alpha=\hbox{tan}^{-1}(v_T/v_R)$, see Fig.~\ref{fig:alpha}. The differential operator $\frac{d}{dt}$ in (\ref{h5}) is defined by the expression
\begin{equation}
\frac{d}{dt} = \frac{\partial}{\partial t} 
+  v \cos\alpha \frac{\partial}{\partial r} 
+ \frac{v \sin\alpha}{r} \frac{\partial}{\partial \phi}.
\end{equation} 
Equations (\ref{hyd2}) are three differential equations for four unknown functions: $T$, $s$, $v$, and $\alpha$. To solve them one has to use also the equation of state, \ie, the relation connecting $T$ and $s$. In our approach, the equation of state is taken into account by the use of the temperature dependent sound velocity 
\begin{equation}
c_s^2(T) = \frac{\partial P}{\partial \epsilon} = 
\frac{s}{T}\frac{\partial T}{\partial s},
\label{cs}
\end{equation}
and by the use of the potential $\Phi(T)$ defined by the differential equation
\begin{equation}
d\Phi=\frac{d\ln T}{c_s}=c_s d\ln s.
\label{phi}
\end{equation}
The form of the function $c_s^2(T)$ used in our calculations is defined and discussed in more detail in Ref. \cite{Chojnacki:2004ec}. Here we only note that the form of our sound velocity function automatically satisfies the condition against the shock formation \cite{Baym:1983sr,Blaizot:1987cc}. The integration of Eq. (\ref{phi}) allows us to express $\Phi$ in terms of the temperature, $\Phi = \Phi_T(T)$, or to express temperature in terms of $\Phi$, $T=T_\Phi(\Phi)$. These two functions may be used to rewrite the hydrodynamic equations in the very concise form. In fact, with the help of the substitutions:
\begin{equation}
v = \tanh \theta, \quad a_\pm = \exp(\Phi \pm \theta),
\label{subst}
\end{equation}
the sum and the difference of the equations (\ref{h3}) and (\ref{h4}) may be rewritten as
\begin{eqnarray}
&& \frac{\partial a_{\pm }}{\partial t}+\frac{\left( v\pm c_{s}\right) }{%
\left( 1\pm c_{s}v\right) }\cos \alpha \frac{\partial a_{\pm }}{\partial r}+%
\frac{\left( v\pm c_{s}\right) }{\left( 1\pm c_{s}v\right) }\frac{\sin
\alpha }{r}\frac{\partial a_{\pm }}{\partial \phi } 
\nonumber \\
&& -\,\frac{c_{s}v}{\left( 1\pm c_{s}v\right) }\left( \sin \alpha \frac{%
\partial \alpha }{\partial r}-\frac{\cos \alpha }{r}\frac{\partial \alpha }{%
\partial \phi }\right) \,a_{\pm }
\nonumber \\
&& 
+ \,\frac{c_{s}}{\left( 1\pm c_{s}v\right) }%
\left[ \frac{1}{t}+\frac{v\cos \alpha }{r}\right] \,a_{\pm } =0,
\label{apmeq}
\end{eqnarray}
while the third equation in (\ref{h5}) is
\begin{eqnarray}
\frac{\partial \alpha }{\partial t} &=&
\frac{\left( 1-v^{2}\right) c_{s}}{v}
\left( \sin \alpha \frac{\partial \Phi }{\partial r}-\frac{\cos \alpha }{r}%
\frac{\partial \Phi }{\partial \phi }\right) 
\nonumber \\
& & 
- v \left( \cos \alpha \frac{\partial \alpha }{\partial r}
+\frac{\sin \alpha }{r}\frac{\partial \alpha }{\partial
\phi }+\frac{\sin \alpha }{r}\right).
\label{alphaeq}
\end{eqnarray}
Equations (\ref{apmeq}) and (\ref{alphaeq}) are three equations for three unknown functions: $a_+(t,r,\phi)$, $a_-(t,r,\phi)$, and $\alpha(t,r,\phi)$. We note, that the velocity $v$ and the potential $\Phi$ are functions of $a_+$ and $a_-$, 
\begin{equation}
v = {a_+ - a_- \over a_+ + a_-}, \quad \Phi = \frac{1}{2} \ln (a_+ a_-).
\label{vPhi}
\end{equation}
Also the sound velocity may be expressed as the function of $a_+$ and $a_-$, however, through a more complicated formula, see Ref. \cite{Chojnacki:2006tv}.
\section{Boundary conditions}
\label{sect:bcond}
\begin{figure}[t]
\begin{center}
\includegraphics[angle=0,width=0.45\textwidth]{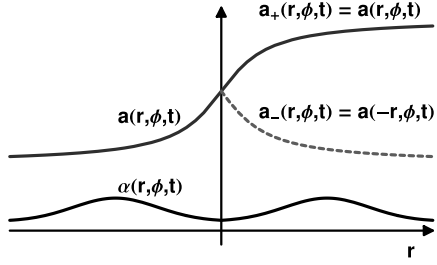}
\end{center}
\caption{Construction of the functions $a_+ (t,r,\phi)$ and $a_-(t,r,\phi)$ with the help of the function $a(t,r,\phi)$. The function $\alpha(t,r,\phi)$ is symmetrically extended to negative values of $r$. }
\label{fig:rsym}
\end{figure}
In this paper we consider the collisions of identical nuclei with atomic number $A$  which collide moving initially along the $z$-axis. The impact parameter ${\bf b}$ points in the $x$-direction. The center of the first nucleus in the transverse plane is placed at ${\bf x}_1 = (x_1,y_1)=(-b/2,0)$, and of the second nucleus at ${\bf x}_2 = (x_2,y_2)=(b/2,0)$. In such a case we require that the magnitude of the flow vanishes at ${\bf x} = (x,y)=(0,0)$. This condition is fulfilled naturally by the ansatz 
\begin{eqletters}
\label{adefwe}
\begin{eqnarray}
a_+(t,r,\phi) &=& a(t,r,\phi),\quad r>0, \label{bc1} \\
a_-(t,r,\phi) &=& a(t,-r,\phi),\quad r<0. \label{bc2}
\end{eqnarray}
\end{eqletters}
We supplement the ansatz (\ref{adefwe}) by the definition of the function $\alpha(t,r,\phi)$ for the negative arguments, see Fig. \ref{fig:rsym},
\begin{eqnarray}
\alpha(t,-r,\phi) &=& \alpha(t,r,\phi), \quad r > 0.
\label{alphadefwe}
\end{eqnarray}
With the help of the definitions (\ref{adefwe}) and (\ref{alphadefwe}), Eqs. (\ref{apmeq}) may be reduced to a single equation for the function $a(t,r,\phi)$,  
\begin{eqnarray}
&& \frac{\partial a}{\partial t}+\frac{\left( v + c_{s}\right) }{
\left( 1 + c_{s}v\right) }\cos \alpha \frac{\partial a}{\partial r}+
\frac{\left( v + c_{s}\right) }{\left( 1 + c_{s}v\right) }\frac{\sin
\alpha }{r}\frac{\partial a}{\partial \phi } 
 \\ \label{aeq}
&& -\,\frac{c_{s}v}{\left( 1 + c_{s}v\right) }\left( \sin \alpha \frac{
\partial \alpha }{\partial r}-\frac{\cos \alpha }{r}\frac{\partial \alpha }{
\partial \phi }\right) \,a + \,\frac{c_{s}}{\left( 1 + c_{s}v\right) }
\left[ \frac{1}{t}+\frac{v\cos \alpha }{r}\right] \,a =0, \nonumber
\end{eqnarray}
Here, similarly to the cylindrically symmetric case, the range of the variable $r$ is extended to negative values. Eq. (\ref{aeq}) should be solved together with Eq. (\ref{alphaeq}), where the range of $r$ may be also extended trivially to negative values (definitions (\ref{adefwe}) and (\ref{alphadefwe}) imply that this equation is in fact invariant under transformation: $r \to -r$). The use of the polar coordinates in the transverse plane requires also that all functions at $\phi=0$ and $\phi=2\pi$ are equal: $a(t,r,0) = a(t,r,2 \pi), \alpha(t,r,0) = \alpha(t,r,2 \pi)$.
\section{Initial conditions}
In the following we assume that the hydrodynamic evolution starts at a typical time $t=t_{0}=1$ fm.  We assume also that the initial temperature profile is connected with the number of participating nucleons
\begin{equation}
T(t_0,x,y) = \hbox{const} \left( \frac{dN_p}{dx\,dy} \right)^{1/3}. 
\label{Tt0}
\end{equation}
The idea to use Eq. (\ref{Tt0}) follows from the assumption that the initially produced entropy density  $\sigma(t_0,x,y)$ is proportional to the number of the nucleons participating in the collision at the position $(x,y)$. Since the considered systems are initially very hot (with the temperature exceeding the critical temperature $T_c$), they may be considered as systems of massless particles, where the entropy density is proportional to the third power of the temperature. In this way we arrive at Eq. (\ref{Tt0}).
The initial form of the functions $v(t,r,\phi)$ and $\alpha(t,r,\phi)$ is
\begin{equation}
v(t_0,r,\phi) = v_{\,0}(r) =  \frac{ H_0 r}{\sqrt{1 + H_0^2 r^2}}, \quad \alpha(t_0,r,\phi) = 0,
\label{initv}
\end{equation}
where $H_0$ is a parameter defining the initial transverse flow formed in the pre-equilibrium stage. In the present calculations we use a very small value $H_0$ = 0.001 fm$^{-1}$. The two initial conditions, Eqs. (\ref{Tt0}) and (\ref{initv}), may be included in the initial form of the function $a(r)$ if we define
\begin{equation}
a(t=t_0,r,\phi) =
\hbox{exp}\left\{\Phi_T \left[ \hbox{const} \,\, \left( \frac{dN_p}{dx\,dy} \right)^{1/3} \right]
\right\} \frac{\sqrt{1 + v^{\,0}(r)}}{\sqrt{1 - v^{\,0}(r)} },
\label{initaT}
\end{equation}
\section{Results}
\begin{figure*}[t!]
\begin{center}
\subfigure{\includegraphics[angle=0,width=\textwidth]{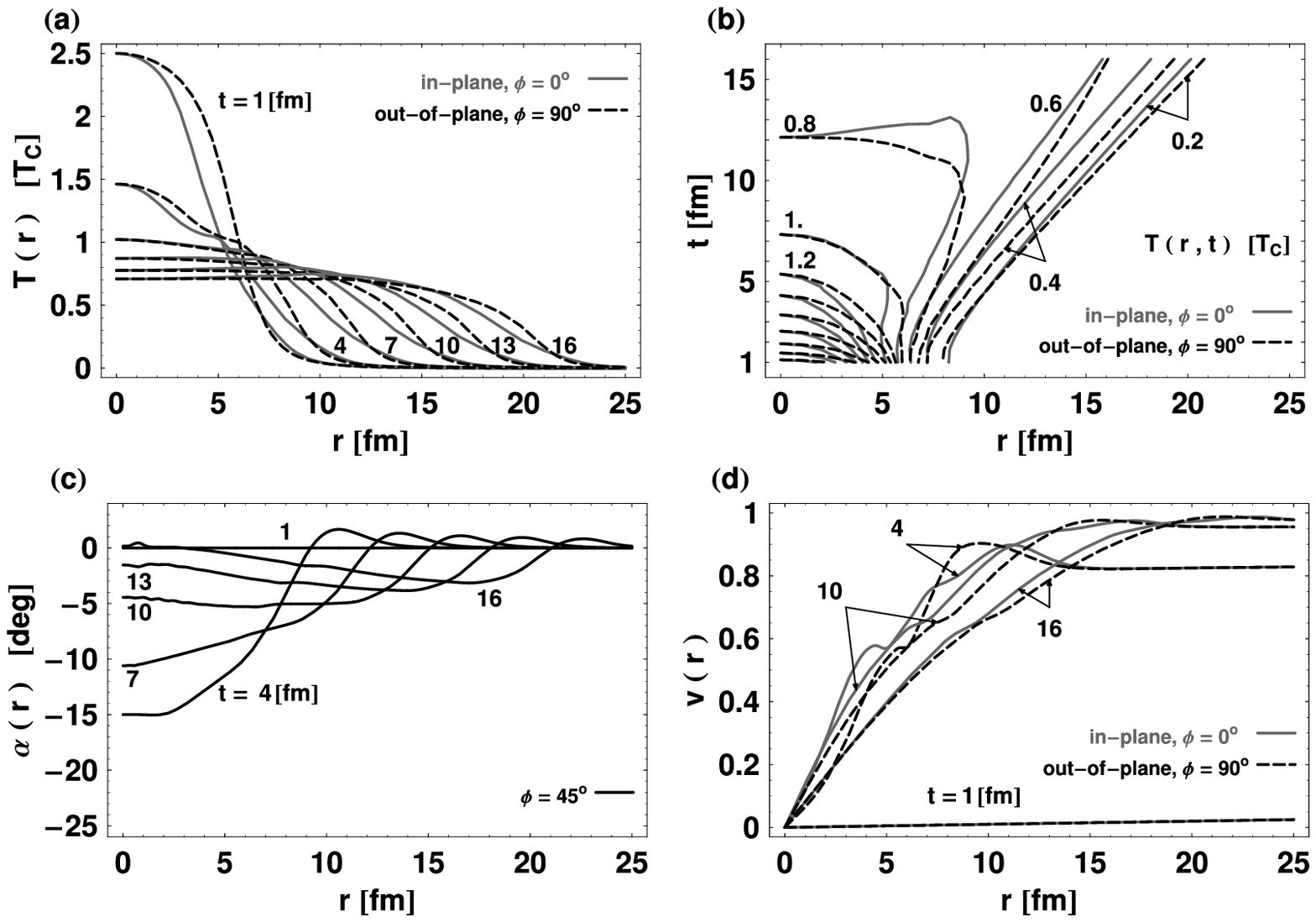}} \\
\subfigure{\includegraphics[angle=0,width=\textwidth]{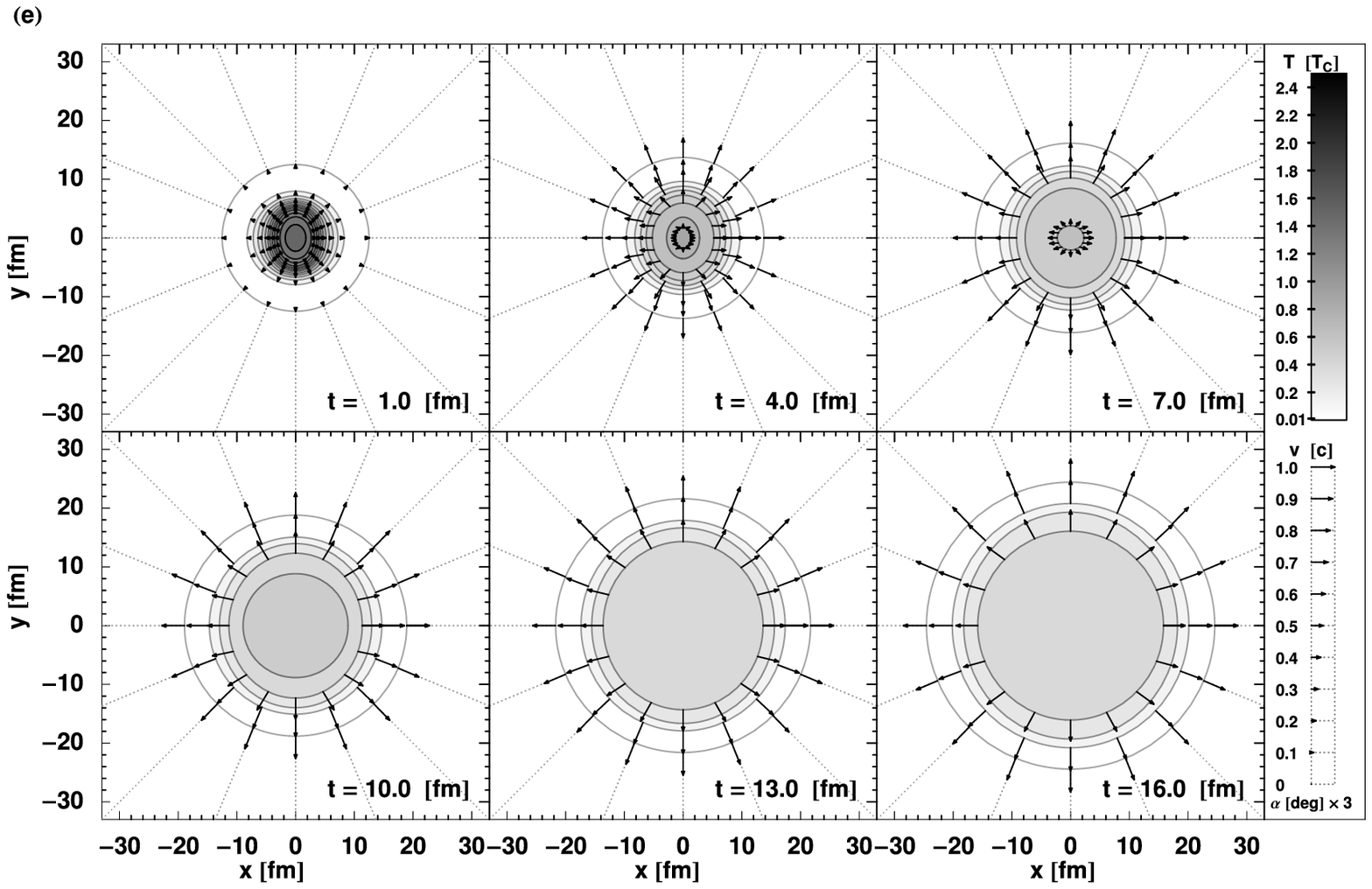}}
\end{center}
\caption{Time development of matter characterized by the initial conditions (\ref{initv}) -- (\ref{initaT}) with $H_0 = 0.001 \hbox{fm}^{-1}$,  $T_0~=~T(t_0,0,0)~=~2.5~T_c$, and $b=7.6$ fm. }
\label{fig:res1}
\end{figure*}

In this section we give an example of our results describing the collisions with the impact parameter $b$ = 7.6 fm, typical for centrality class $c=0-80\%$ \cite{Broniowski:2001ei} \footnote{In the paper \cite{Chojnacki:2006tv} we considered the centrality classes: 0-20\%, 20-40\%, 40-60\%}. In this case the initial central temperature $T_0$ equals 2.5 $T_c$, where $T_c$ is the critical temperature, $T_0 = T(t_0,0,0)$. The part a) of Fig. \ref{fig:res1} shows the temperature profiles for different values of time: $t = 1, 4, 7, 10, 13, 16$ fm. The solid lines represent the temperature profiles along the $x$-axis ($\phi=0$), while the dashed lines represent the profiles along the $y$-axis ($\phi=\pi/2$). One can observe that during the whole considered evolution time, 1 fm $ < t < $ 16 fm, the $y$-extension of the system remains larger than the $x$-extension.  However, the relative magnitude of this effect decreases with time, indicating that the cylindrical symmetry is gradually restored as the time increases. In this part of Fig. \ref{fig:res1} one can also notice the effect of the phase transition; initially the system cools down rather rapidly, later the cooling down is delayed and the main visible effect is the increase of the volume of the system. We note that this behavior is related to the sudden decrease of the sound velocity in the region $T \approx T_c$. The part b) of Fig. \ref{fig:res1} shows the isotherms in the $t-r$ space, again for $\phi=0$ (solid lines) and $\phi=\pi/2$ (dashed lines). The pairs of isotherms indicate different values of the temperature. They start at $T = 1.8 \,T_c$ and go down to $T = 0.2 \,T_c$, with a step of $0.2 \,T_c$. It is interesting to observe that the solid and dashed lines cross each other. This effect means that the central (relatively hotter) part of the system acquires a pumpkin-like shape during the evolution of the system. Such pumpkin-like regions, however, shrink and disappear during further expansion. In the part c) of Fig. \ref{fig:res1} the profiles of the function $\alpha(t,r,\phi)$ are shown for $t = 1, 4, 7, 10, 13, 16$ fm and $\phi = \pi/4$. For $\phi=0$ and $\phi=\pi/2$ the function $\alpha(t,r,\phi)$ vanishes due to the symmetry reasons. In the first and third quadrant the values of $\alpha$ are predominantly negative, while in the second and fourth quadrant the values of $\alpha$ are positive. This behavior characterizes the direction of the flow which has the tendency to change the initial almond shape into a cylindrically symmetric shape. In the part d) the velocity profiles are shown, again for $t$ = 1, 4, 10, 16 fm. Similarly to the previously discussed figures, the solid lines are the profiles for $\phi=0$ ($x$-direction), whereas the dashed lines are the profiles for $\phi=\pi/2$ ($y$-direction). One can observe that the magnitude in the $x$-direction is larger than the magnitude in the $y$-direction, which is an expected hydrodynamic behavior caused by larger pressure gradients in the $x$-direction. Exactly this effect is responsible for the observed azimuthal asymmetry of the transverse-momentum spectra, quantified by the values of the $v_2$ coefficient. Finally, in the part e) we show the contour lines of the temperature, again for different values of time. These plots visualize the time development of the system. The arrows describe the magnitude and the direction of the flow (for better recognition the angle $\alpha$ is magnified by a factor of 3). 
\section{Summary}
In this paper we presented a new and concise treatment of the boost-invariant and cylindrically asymmetric relativistic hydrodynamic equations. The presented formalism is a direct generalization of the approach introduced by Baym et al. in Ref. \cite{Baym:1983sr}. In the studied case, the symmetry of the problem allows us to rewrite the hydrodynamic equations as only two coupled partial differential equations, (\ref{alphaeq}) and (\ref{aeq}), which automatically lead to the fulfillment of the requested boundary conditions for the velocity and the temperature at the center of the system. The effects of the phase transition are included in this scheme by the use of the temperature dependent sound velocity. The presented results of the hydrodynamic calculations, supplemented with the appropriate freeze-out model (\eg, the single-freeze-out model of Refs. \cite{Broniowski:2001we,Broniowski:2001uk}), may be used to calculate different physical observables. To achieve this task, in the closest future we intend to combine our hydrodynamic approach with the statistical Monte-Carlo model {\tt THERMINATOR} \cite{Kisiel:2005hn}.
\end{document}